\documentstyle[pre,preprint,aps]{revtex}
\tightenlines
\begin{document}
\draft
\title{
Slave-boson mean-field theory of spin- and orbital-ordered states 
in the degenerate Hubbard model\footnote{To be submitted to 
a special issue of "Foundation of Physics" celebrating the 75-th birthday of 
Martin Gutzwiller}
} 
\author{
Hideo Hasegawa\footnote{E-mail: hasegawa@u-gakugei.ac.jp}
}
\address{
Department of Physics, Tokyo Gakugei University,
Koganei, Tokyo 184, Japan.\\
}
\maketitle
\begin{abstract}
The mean-field theory 
with the use of the slave-boson functional method 
is generalized to take account of the spin- and orbital-ordered
state in the doubly degenerate Hubbard model.
Some numerical calculations are presented of 
the antiferromagnetic orbital-ordered state
in the half-filled simple-cubic model.
The orbital order in the present theory is much
reduced compared with that in the Hartree-Fock approximation
because of the large orbital fluctuations. 
From a comparison of the ground-state energy,
the antiferromagnetic orbital state is shown to 
be unstable against the antiferromagnetic spin state, 
although the situation becomes reversed
when the exchange interaction is {\it negative}.

\end{abstract}

%
\newpage

\begin{center}
{\bf I. INTRODUCTION}
\end{center}
  
  In his seminal paper, Gutzwiller\cite{Gutzwiller63} proposed  
a variational approach, employing the projected wave function to 
take into account the effect of correlated electrons.
Since the exact evaluation of the ground-state energy is difficult,
he introduced an additional approximation
which is now called the Gutzwiller approximation (GA).
A validity of the GA was studied by Monte-Carlo method
for finite-size clusters \cite{Kaplan82}\cite{Yokoyama87}.  
It is realized that
the GA becomes a better approximation for higher-dimensional
systems, and it is exact in the limit of infinite dimension.
The GA has been widely adopted in many area including
metal-insulator transition (MIT) \cite{Brinkman70}, 
magnetism and high-$T_c$ superconductors 
(for a review of the GA of SHM see Ref. \cite{Gebhard97}).

Kotliar and Ruckenstein \cite{Kotliar86} developed the slave-boson
mean-field theory by using the saddle-point approximation 
to the slave-boson functional-integral method.
They successfully derived the antiferromagnetic solution besides the 
paramagnetic solution of the GA.
It is shown that the GA and the slave-boson
mean-field theory is equivalent.

Most of the theoretical studies based on the GA 
have been made for the single-band
Hubbard model (SHM) for its simplicity.
Actual systems, however, inevitably have the orbital degeneracy.
It is necessary to investigate the role of the orbital degeneracy
and the effect of Hund-rule coupling due to the exchange 
interaction for a better understanding on strongly 
correlated systems.

  The first attempt to include the orbital degeneracy within the GA
was made by Chao and Gutzwiller \cite{Chao71}.
In the last few years the Hubbard model with orbital degeneracy
has been extensively studied by using not only the
GA-type theories \cite{Lu94}-\cite{Fresard97} but also
the dynamical mean-field approximation 
\cite{Kotliar96},\cite{Rozenberg96},
and Monte-Carlo simulations \cite{Gill87}-\cite{Motome98}.
The original GA proposed by Chao and Gutzwiller \cite{Chao71}
was reformulated in Refs. \cite{Lu94}-\cite{Bunemann98}.
Lu \cite{Lu94} obtained the analytical expression
of the critical interaction for the MIT.
Okabe \cite{Okabe96} proposed a sophisticated method in calculating
te band-narrowing factor which is a difficult part
in applying the GA to the degenerate band model.
The first-order transition with an introduction
of the exchange interaction was discussed in Ref. \cite{Bunemann97}.
B\"{u}nemann, Weber and Gebhard \cite{Bunemann98}
made a detailed study by using the generalized interactions.

The present author \cite{Hasegawa97a} proposed
a slave-boson 
mean-field theory for the degenerate Hubbard model,
by employing the slave-boson method proposed 
by Dorin and Schlottman \cite{Dorin93} for 
the Anderson lattice model.
Fr\'{e}sard and Kotliar\cite{Fresard97} 
adopted an alternative slave-boson
functional integral method.
The slave-boson mean-field theory of Hasegawa 
\cite{Hasegawa97a} and Fr\'{e}sard and Kotliar\cite{Fresard97} 
is the simple generalization of
the Kotliar-Ruckenstein theory \cite{Kotliar86} for the SHM 
to that for the degenerate Hubbard model, and
it is again equivalent to the GA \cite{Lu94}-\cite{Bunemann98}.
The MIT in the paramagnetic state
in the doubly degenerate Hubbard model (DHM)
is discussed \cite{Hasegawa97c}\cite{Fresard97}. 
Subsequently the slave-boson theory was extended and
applied to the antiferromagnetic (spin)
state of the half-filled DHM \cite{Hasegawa97b}.

  One of the advantages of the slave-boson
mean-filed theory to the GA is
that it has the wider applicability than
the GA. By using the Green-function method,
for example, the slave-boson mean-field theory
can easily deal with the 
antiferromagnetic state in the SHM 
\cite{Kotliar86},\cite{Hasegawa89}
and in the doubly 
degenerate Hubbard model (DHM) \cite{Hasegawa97b}.
This suggests that we can discuss
a system with more complicated orderings
with the slave-boson mean-field theory.
It is the purpose of the present paper to generalize
our theory to include the orbital ordering besides
spin ordering.

  The paper is organized as follows:  In the next Sec. II,
we  present a basic formulation of our slave-boson saddle-point 
approximation to deal with the spin- and orbital-ordered states.
Since calculations for the paramagnetic and antiferromagnetic
states have been reported in our previous papers
\cite{Hasegawa97a}-\cite{Hasegawa97c}, we present,
in Sec. III, some calculated results for the orbital-ordered state in the 
half-filled DHM. 
Section IV is devoted to conclusion and supplementary discussion. 

\vspace{0.5cm}
 
\begin{center}
{\bf II. FORMULATION}
\end{center}

\noindent
{\it 2.1 Basic Equations}
    
   We adopt the Hubbard model with the doubly orbital degeneracy
whose Hamiltonian is given by
\begin{eqnarray}
H & = & \sum_{\sigma} \sum_{i j} \sum_{\tau \tau'} 
t^{\tau\tau'}_{ij} 
c^\dagger_{i\tau \sigma} c_{j\tau' \sigma}
+ \frac{1}{2} \sum_i \sum_{(\tau,\sigma) \neq (\tau',\sigma')}
U_{\tau\tau'}^{\sigma\sigma'}
c^\dagger_{i\tau \sigma} c_{i\tau \sigma}
c^\dagger_{i\tau' \sigma'} c_{i\tau' \sigma'} \nonumber \\
& & -  \sum_{\sigma} \sum_{i } \sum_{\tau} 
(\sigma h_i + \tau g_i) \;
c^\dagger_{i\tau \sigma} c_{i\tau' \sigma},
\end{eqnarray}
\noindent
where $c_{i\tau\sigma}$ is an annihilation operator of an electron 
with an orbital index $\tau (= \pm 1)$ 
and spin $\sigma \: (= \pm 1)$ on the lattice site $i$.
The first term expresses electron hopping, which is assumed to 
be allowed only between the same subband:
$t_{ij}^{\tau\tau'} = t_{ij} \delta_{\tau\tau'}$, for a simplicity.
The on-site interaction, $U_{\tau\tau'}^{\sigma \sigma'}$, 
in the second term is assumed to be given by
\begin{eqnarray}
U_{\tau\tau'}^{\sigma\sigma'} &=& U_0 = U   \;\;\;\;\;\;\;\;\;\;
\mbox{for $\tau = \tau', \sigma \neq \sigma'$},  \\
&=& U_1 = U - 2J  \;\; 
\mbox{for $\tau \neq \tau', \sigma \neq \sigma'$},  \\
&=& U_2 = U - 3 J  \;\; 
\mbox{for $\tau \neq \tau', \sigma = \sigma'$}, 
\end{eqnarray}
where $U$ and $J$ are Coulomb and exchange interactions, respectively.
In the third term of Eq.(1), $h_i$ and $g_i$ are the magnetic 
and crystal fields, respectively, at the $i$ site.

  In order to discuss the DHM
with the slave-boson functional integral method, 
we introduce, for a given site $i$,
sixteen slave bosons which are classified 
into the following five categories:

\noindent
(1) $e_i$ for the empty state;

\noindent
(2) $p_{i\tau}$ for the singly occupied state with a $\sigma$-spin
electron in the $\tau$ band;

\noindent
(3) $d_{i0\tau}$ for the doubly occupied state with a pair of 
up- and down-spin electrons in the $\tau$ band;
$d_{i1\tau}$ for that with an up-spin electron
in the $\tau$ band and a down-spin electron in the $-\tau$ band;
$d_{i2\sigma}$ for that with $\sigma$-spin
electrons in the two subbands;

\noindent
(4) $t_{i\tau\sigma}$ for the triply occupied state with a pair of 
up- and down-spin electrons in the $\tau$ band and an extra
$\sigma$-spin electron in the $-\tau$ band; and

\noindent
(5) $f_i$ for the fully occupied state. 

\noindent
These slave bosons obey the constraint given by

\begin{equation}
e_i + \sum_{i\tau\sigma} p_{i\tau\sigma} 
+ \sum_{\tau} (d_{i0\tau} + d_{i1\tau})
+ \sum_{\sigma} d_{i2\sigma}  
+  \sum_{\sigma} \sum_{\tau} t_{i\tau\sigma} + f_i = 1,
\end{equation}
and the equivalence between fermion and boson operators is
given by
\begin{equation}
c^{\dagger}_{i\tau\sigma}c_{i\tau\sigma} 
= p_{i\tau\sigma} + d_{i0\tau} + d_{i1\:\sigma\tau}
+ d_{i2\sigma} + t_{i\tau\sigma} 
+ t_{i\tau-\sigma} + t_{i-\tau\sigma} + f_i.
\end{equation}

By incorporating  the conditions given 
by Eqs.(5) and (6) with the Lagrange
multipliers, $\lambda_i^{(1)}$ and $\lambda^{(2)}_{i\tau\sigma}$,
we get the partition function given as
a functional integral over the coherent state of Fermi and
Bose fields \cite{Hasegawa97a}:
\begin{equation}
Z = 
\int \! De\int \! Dp
\int \! Dd_{0}\int \! Dd_{1}\int \! Dd_{2}
\int \! Dt \int \! Df 
\int \! D\lambda^{(1)} \int \!\!D\lambda^{(2)}
\:\: {\rm exp} \left[- \int_{0}^{\beta}
\left( L_f(t) + L_b(t) \right)  \right],
\end{equation}
with
\begin{equation}
L_f(t) = \sum_{i j} \sum_{\sigma} \sum_{\tau} 
z_{i\tau\sigma} \;t_{ij}\;z_{j\tau\sigma}\; 
c^\dagger_{i\tau \sigma} c_{j\tau \sigma}
+ \sum_i \sum_{\sigma} \sum_\tau  
(\frac{\partial}{\partial t} + \lambda_{i\tau\sigma}^{(2)}
- \sigma\; h_i - \tau\; g_i) \;
c^\dagger_{i \tau\sigma} c_{i\tau \sigma}.
\end{equation}
\begin{eqnarray}
L_b(t) & =& \sum_i \left[ \sum_{\tau} (U_0 d_{i0\tau} +  U_1 d_{i1\tau}) 
+ U_2 \sum_{\sigma}\;d_{i2\sigma}
+ (U_0 + U_1 + U_2 ) \left( (\sum_{\sigma} \sum_{\tau}
t_{i\tau\sigma}) + 2\:f_i \right) \right]  \nonumber \\
& + & \sum_i (\frac{\partial}{\partial t} + \lambda_i^{(1)}) 
\left[ e_i + \sum_{i\tau\sigma} p_{i\tau\sigma} 
+ \sum_{\tau} (d_{i0\tau} + d_{i1\tau})
+ \sum_{\sigma} d_{i2\sigma}  
+  \sum_{\sigma} \sum_{\tau} t_{i\tau\sigma} + f_i - 1 \right] 
\nonumber \\
& - & \sum_i \sum_{\sigma} \sum_{\tau} \lambda_{i\tau\sigma}^{(2)}
\left[ p_{i\tau\sigma} + d_{i0\tau} + d_{i1\:\sigma\tau}
+ d_{i2\sigma} + t_{i\tau\sigma} 
+ t_{i\tau-\sigma} + t_{i-\tau\sigma} + f_i \right].
\end{eqnarray}
Here 
$Dp  = \Pi_i \Pi_{\sigma} \Pi_{\tau} \; 
d\;p_{i\tau\sigma}$ {\it et al.},
$\beta$ is the inverse temperature,
$L_f$ and $L_b$ denote the terms relevant to the Fermi
and Bose fields, respectively, 
and $z_{i\tau\sigma}$ is given by
\begin{eqnarray}
z_{i\tau\sigma} & = & 
(n_{i\tau\sigma})^{-1/2} \:
( \surd \overline{e_i p_{i\tau\:\sigma}} 
+ \surd \overline{p_{i\tau\:-\sigma} d_{i0\:\tau}}
+ \surd \overline{p_{i-\tau\:-\sigma}d_{i1\:\sigma\tau}}
+ \surd \overline{p_{i-\tau\:\sigma}\:d_{i2\:\sigma}} \nonumber \\ 
 & & +  \surd \overline{d_{i0\:-\tau}\:t_{i-\tau\:\sigma}}   
+ \surd \overline{d_{i1\:-\sigma\tau}\:t_{i\tau\:\sigma}}
+ \surd \overline{d_{i2\:-\sigma}\:t_{i\tau\:-\sigma}}
+ \surd \overline{d_{i-\tau\:-\sigma}\:f_i} )
\; (1 - n_{i\tau\sigma})^{-1/2}.
\end{eqnarray}
where $n_{i\tau\sigma} 
\;=\;< c^{\dagger}_{i\tau\sigma} c_{i\tau\sigma} >$ and
the bracket denotes the expectation value. 

In discussing the spin- and/or orbital-orderded states, 
we make the following change of variables for $n_{i\tau\sigma}$:
\begin{equation}
N_i = \sum_{\sigma} \sum_{\tau} n_{i\tau\sigma},
\end{equation}
\begin{equation}
M_i = \sum_{\sigma} \sum_{\tau} \sigma \; n_{i\tau\sigma},
\end{equation}
\begin{equation}
O_i = \sum_{\sigma} \sum_{\tau} \tau \; n_{i\tau\sigma},
\end{equation}
\begin{equation}
P_i = \sum_{\sigma} \sum_{\tau} \sigma\tau \;n_{i\tau\sigma},
\end{equation}
where $N_i$, $M_i$, $O_i$ and $P_i$ denote operators relevant to
the number of electrons,
the  magnetic moment, the orbital order and the orbital spin 
polarization, respectively, at a given $i$ site.
Similarly, we make the change of variables for 
$\lambda_{i\tau\sigma}^{(2)}$ as follows: 
\begin{equation}
\nu_i = (1/4)\;\sum_{\sigma} \sum_{\tau}  
\lambda_{i\tau\sigma}^{(2)},
\end{equation}
\begin{equation}
\xi_i = (- 1/4)\;\sum_{\sigma} \sum_{\tau} \sigma \;
\lambda_{i\tau\sigma}^{(2)},
\end{equation}
\begin{equation}
\phi_i = (- 1/4)\;\sum_{\sigma} \sum_{\tau} \tau \;
\lambda_{i\tau\sigma}^{(2)},
\end{equation}
\begin{equation}
\eta_i = (- 1/4)\;\sum_{\sigma} \sum_{\tau} \sigma\tau \;
\lambda_{i\tau\sigma}^{(2)},
\end{equation}
where $\nu_i$, $\xi_i$, $\phi_i$ and $\eta_i$ 
stand for conjugate fields of $N_i$, $M_i$, $O_i$ and $P_i$, 
respectively. Then $n_{i\tau\sigma}$ 
and $\lambda_{i\tau\sigma}^{(2)}$ are expressed by
\begin{equation}
n_{i\tau\sigma} = (1/4) \; (N_i + \sigma \;M_i + \tau \;O_i 
+ \sigma \tau \; P_i)
\end{equation}
\begin{equation}
\lambda^{(2)}_{i\tau\sigma} 
= (\nu_i - \sigma \;\xi_i - \tau \;\phi_i 
- \sigma \tau \; \eta_i)
\end{equation}

Substituting Eqs.(19) and (20) to Eqs.(7)-(9) 
with some manipulations, we get the partition function 
within the static approximation, given by 
%
\begin{eqnarray}
Z & = &
\int \! DN \int \! DM \int \! DO \int \! DP
\int  \! D\nu \int  \! D\xi \int  \! D\phi \int  \! D\eta  
\nonumber \\
& & \mbox{     } \;\;\;\;\;\;\;\;\;\;\;\;\; \times
\int \! Dd_{0}\int \! Dd_{1}\int \! Dd_{2} 
\int \! Dt \int \! Df 
\:\: {\rm exp} (- \beta  \Phi),
\end{eqnarray}
%
%
with
\begin{equation}
\Phi = \Phi_0+\Phi_1+\Phi_2,
\end{equation}
\begin{equation}
\Phi_0 = \sum_i \left[ \sum_{\tau} (U_0 d_{i0\tau} +  U_1 d_{i1\tau}) 
+ U_2 \sum_{\sigma}\;d_{i2\sigma}
+ (U_0 + U_1 + U_2 ) \left( (\sum_{\sigma} \sum_{\tau}
t_{i\tau\sigma}) + 2\:f_i \right) \right],
\end{equation}
\begin{equation}
\Phi_1 = \sum_{i} 
\left[ (\varepsilon_{F} - \nu_i)N_i 
+  \xi_i M_i + \phi_i O_i + \eta_i P_i \right],
\end{equation}
\begin{equation}
\Phi_2 = \int d\varepsilon \; f(\varepsilon)\;  (- 1/\pi) \; 
{\rm Im} \:\:
{\rm Tr} \:\: {\rm ln} \:  G(\varepsilon). 
\end{equation}
Here $f(\varepsilon)$ is the 
Fermi-distribution function, $\varepsilon_F$  the
Fermi energy and $G(\varepsilon)$ is the one-particle Green function
given by 
\begin{equation}
G(\varepsilon) = (\varepsilon - H_{\rm eff})^{-1},
\end{equation}
where the effective slave-boson Hamiltonian, $H_{\rm eff}$, is given by
\begin{equation}
H_{\rm eff} = 
\sum_{i j} \sum_{\sigma} \sum_{\tau} 
q_{\tau\sigma}^{ij}
t_{ij} 
c^\dagger_{i\tau \sigma} c_{j\tau \sigma}
+ \sum_i \sum_{\sigma} \sum_\tau  
\left[ \nu_{i} - \sigma (h_i + \xi_{i})  
- \tau (g_i +  \phi_i) 
- \sigma\tau\: \eta_i \right] \:
c^\dagger_{i \tau\sigma} c_{i\tau \sigma}.
\end{equation}
and the band-narrowing factor,
$q_{\tau\sigma}^{ij}$, is given by
\begin{equation}
q_{\tau\sigma}^{ij} = z_{i\tau\sigma} \; z_{j\tau\sigma},
\end{equation}
with
\begin{equation}
z_{i\tau\sigma} = C_{i\tau\sigma}/D_{i\tau\sigma},
\end{equation}
\begin{eqnarray}
C_{i\tau\sigma} & = & 16 
( \surd \overline{e_i p_{i\tau\:\sigma}} 
+ \surd \overline{p_{i\tau\:-\sigma} d_{i0\:\tau}}
+ \surd \overline{p_{i-\tau\:-\sigma}d_{i1\:\sigma\tau}}
+ \surd \overline{p_{i-\tau\:\sigma}\:d_{i2\:\sigma}} \nonumber \\ 
 & & +  \surd \overline{d_{i0\:-\tau}\:t_{i-\tau\:\sigma}}   
+ \surd \overline{d_{i1\:-\sigma\tau}\:t_{i\tau\:\sigma}}
+ \surd \overline{d_{i2\:-\sigma}\:t_{i\tau\:-\sigma}}
+ \surd \overline{d_{i-\tau\:-\sigma}\:f_i} ),
\end{eqnarray}
\begin{eqnarray}
D_{i\tau\sigma} = (N_i + \sigma M_i + \tau O_i 
+ \tau\:\sigma P_i )^{1/2} \:
(4 - N_i - \sigma M_i - \tau O_i - \tau\:\sigma P_i )^{1/2}  .
\end{eqnarray}
\begin{equation}
e_i = 1 - N_i + \sum_{\tau} (d_{i0\tau} + d_{i1\tau})
+ \sum_{\sigma} d_{i2\sigma}  
+ 2 \sum_{\sigma} \sum_{\tau} t_{i\tau\sigma} + 3f,
\end{equation}
\begin{equation}
p_{i\tau\sigma} 
=  (1/4) ( N_i + \sigma\;M_i + \tau\;O_i + \sigma\tau P_i )
-(d_{i0\tau} + d_{i1\;\sigma\tau} + d_{i2\sigma}
+ \sum_{\sigma} t_{i\tau\sigma} + t_{i-\tau\sigma} + f_i)
\end{equation}

Since the effective transfer integral in Eq.(27)
is expressed as a product form: \\
$z_{i\tau\sigma} t_{ij} z_{j\tau\sigma}$,
we can express $G(\varepsilon)$ and $\Phi_2$ in terms of 
the locator defined by
\begin{equation}
X_{i\tau\sigma}(\varepsilon) 
= \left[  \varepsilon - \nu_{i} + \sigma (h_i + \xi_{i} )
+ \tau (g + \phi_i ) + \sigma\tau \eta_i \right]
/r_{i\tau\sigma},
\end{equation}
where 
\begin{equation}
r_{i\tau\sigma} = (z_{i\tau\sigma})^2,
\end{equation}
explicit expressions for $\Phi_2$ being given shortly 
[Eqs. (61),(62)].

By evaluating the saddle-point value of $\Phi$,
we get the slave-boson mean-field free energy given by

\begin{eqnarray}
F & =& \sum_i \left[ \sum_{\tau} (U_0 d_{i0\tau} +  U_1 d_{i1\tau}) 
+ U_2 \sum_{\sigma}\;d_{i2\sigma}
+ (U_0 + U_1 + U_2 ) \left( (\sum_{\sigma} \sum_{\tau}
t_{i\tau\sigma}) + 2\:f_i \right) \right]  \nonumber  \\
& & + \sum_{i} 
\left[ (\varepsilon_{F} - \nu_i)N_i 
+  \xi_i M_i + \phi_i O_i + \eta_i P_i \right]  \nonumber  \\
& & + \int d\varepsilon \; f(\varepsilon)\;  (- 1/\pi) \; 
{\rm Im} \:\:
{\rm Tr} \:\: {\rm ln} \:  G(\varepsilon).   
\end{eqnarray}
It is noted that for a given set of parameters of
$U$, $J$, $T$ and $N$, the
total number of electrons (per site)  given by
\begin{equation}
N =  (1/N_0) \sum_i N_i
= (1/N_0) \sum_i \sum_{\sigma} \sum_{\tau} n_{i\tau\sigma}.
\end{equation}
where $N_0$ is the number of lattice sites. The optinum values of
$N_i$, $M_i$, $O_i$, $P_i$, $\nu_i$, $\xi_i$, $\phi_i$, $\eta_i$,
$d_{i0\tau}$, $d_{i1\tau}$, $d_{i2\sigma}$, 
$t_{i\tau\sigma}$,  $f_i$ and $\varepsilon_{\rm F}$ in Eq.(36)
are determined by the self-consistent equations 
given by Eqs.(11)-(14) and the following equations:
\begin{equation} 
\varepsilon_F - \nu_{i} 
+ \sum_\sigma \sum_{\tau} R_{i\tau\sigma} 
(\partial r_{i\tau\sigma} /\partial N_i) = 0,
\end{equation}
\begin{equation} 
\xi_{i} + \sum_\sigma \sum_{\tau} R_{i\tau\sigma} 
(\partial r_{i\tau\sigma} /\partial M_i) = 0,
\end{equation}
\begin{equation} 
\phi_{i} + \sum_\sigma \sum_{\tau} R_{i\tau\sigma} 
(\partial r_{i\tau\sigma} /\partial O_i) = 0,
\end{equation}
\begin{equation} 
\eta_{i} + \sum_\sigma \sum_{\tau} R_{i\tau\sigma} 
(\partial r_{i\tau\sigma} /\partial P_i) = 0,
\end{equation}
\begin{equation}
U_0 + \sum_{\sigma} \sum_{\tau'} R_{i\tau'\sigma} 
(\partial r_{i\tau'\sigma}/\partial d_{i0\tau})  = 0,
\end{equation}
\begin{equation}
U_1 + \sum_{\sigma} \sum_{\tau'} R_{i\tau'\sigma} 
(\partial r_{i\tau'\sigma}/\partial d_{i1\tau})  = 0,
\end{equation}
\begin{equation}
U_2 + \sum_{\sigma'} \sum_{\tau} R_{i\tau\sigma'} 
(\partial r_{i\tau\sigma'}/\partial d_{i2\sigma})  = 0,
\end{equation}
\begin{equation}
U_0 + U_1 + U_2 
+  \sum_{\sigma'} \sum_{\tau'} R_{i\tau'\sigma'} 
(\partial r_{i\tau'\sigma'}/\partial t_{i\tau\sigma})  = 0,
\end{equation}
\begin{equation}
2 (U_0 + U_1 + U_2) 
+ \sum_{\sigma} \sum_{\tau} R_{i\tau\sigma} 
(\partial r_{i\tau\sigma}/\partial f_i)  = 0,
\end{equation}
with
\begin{equation}
R_{i\tau\sigma} = \partial \Phi_2/\partial r_{i\tau\sigma}.
\end{equation}

\begin{equation}
n_{i\tau\sigma} = \int d\varepsilon \; f(\varepsilon)\;  
(-1/\pi) \; Im \:\: 
<i\tau\sigma \mid G(\varepsilon) \mid i\tau\sigma >,
\end{equation}
where $r_{i\tau\sigma}$ is given by Eq.(35), and
the bra $<i\tau\sigma \mid$ and ket $\mid i\tau\sigma>$ 
denote the $\sigma$-spin state
in the $\tau$ subband at the $i$ site.

\vspace{0.5cm}
\noindent
{\it 2.2 Spin- and Orbital-Ordered States}

The expression for $\Phi_2$ given by Eq.(25) 
(and then $R_{i\tau\sigma}$ in Eq.(47)) depends on 
$G(\varepsilon)$ given by Eqs.(26) and (27)
which is specified by
the electronic structure and by the spin- and orbital-ordered states
to be investigated.
In order to discuss the spin- and/or orbital-ordered states 
in the paramagnetic, ferromagnetic or antiferromagnetic state 
on the same footing, we divide the crystal 
into two sublattices, A and B.
We assume that for the AF wave vector $Q$, the relation:
$\varepsilon_{k+Q} = - \varepsilon_k$ holds where
$\varepsilon_k$ is the Fourier transform of the transfer integral,
$t_{ij}$.

The locator given by Eq.(34) at the  $i$ ($j$) site
belonging to the A (B) lattice, is given by
\begin{eqnarray}
X_{i\tau\sigma}(\varepsilon) 
\equiv X_{A\tau\sigma}(\varepsilon) 
& = & \left[  \varepsilon - \nu_A + \sigma (h_A + \xi_A)
+ \tau (g_A + \phi_A) 
+ \sigma\tau \eta_A \right] /r_{A\tau\sigma} 
\;\;\;\;\;\;\;\mbox{$(i \in A)$},  \\
X_{j\tau\sigma}(\varepsilon) 
\equiv X_{B\tau\sigma}(\varepsilon) 
& = & \left[  \varepsilon - \nu_B + \sigma (h_B + \xi_B)
+ \tau (g_B + \phi_B) 
+ \sigma\tau \eta_B \right] /r_{B\tau\sigma} 
\;\;\;\;\;\;\;\mbox{$(j \in B)$}. 
\end{eqnarray}
When the spin and orbital orderings in the A and B sublattices
have the simple symmetry relation such as
$\nu_A = \nu_B = \nu$, $\xi_A = \pm \; \xi_B = \xi$ 
and $\eta_A = \pm \; \eta_B = \eta$ 
with $h_A = h_B = g_A = b_B= 0$, 
Eqs.(49) and (50) become
\begin{eqnarray}
X_{A\tau\sigma}(\varepsilon) 
& =& ( \varepsilon - \nu + \sigma  \xi
+ \tau \phi + \sigma\tau \eta )  /r_{A\tau\sigma} 
\;\;\;\;\;\;\;\;   \\
X_{B\tau\sigma}(\varepsilon) 
& =& ( \varepsilon - \nu + \sigma  \xi
+ \tau \phi + \sigma\tau \eta )  /r_{B\tau\sigma} 
\;\;\;\;\;\;\;\; \mbox{for F-f state},  \\
& = & ( \varepsilon - \nu - \sigma \xi
+ \tau \phi - \sigma\tau \eta ) /r_{B\tau\sigma} 
\;\;\;\;\;\;\;\; \mbox{for AF-f state}, \\
& = & ( \varepsilon - \nu + \sigma  \xi
- \tau \phi - \sigma\tau \eta ) /r_{B\tau\sigma} 
\;\;\;\;\;\;\;\; \mbox{for F-af state}, \\
& = & ( \varepsilon - \nu - \sigma \xi
- \tau \phi + \sigma\tau \eta ) /r_{B\tau\sigma} 
\;\;\;\;\;\;\;\; \mbox{for AF-af state}, \;\;\;\;\;
\end{eqnarray}
with $r_{A\tau\sigma}$ and $r_{B\tau\sigma}$ given by
\begin{eqnarray}
r_{i\tau\sigma} \equiv r_{A\tau\sigma} 
& = & (z_{\tau\sigma})^2,
\;\;\;\;\;\;\;\;\;\;\;\;\;\;\;\;\;\;\;\;\;\;\;\;\;\;
\mbox{$(i \in A)$} \\
%
r_{j\tau\sigma} \equiv r_{B\tau\sigma} 
& = & r_{A\tau\sigma}  
\;\;\;\;\;\;\;\;\;\;\;\; \mbox{for F-f state},  \\
& = & r_{A\tau-\sigma}  
\;\;\;\;\;\;\;\;\;\;\; \mbox{for AF-f state}, \\
& = & r_{A-\tau\sigma}  
\;\;\;\;\;\;\;\;\;\;\; \mbox{for F-af state}, \\
& = & r_{A-\tau-\sigma}  
\;\;\;\;\;\;\;\;\;\; \mbox{for AF-af state}. 
\;\;\;\;\;\mbox{$(j \in B)$}
\end{eqnarray}
In Eqs.(51)-(60)
"AF-f state", for example, expresses the state in which 
spin and orbital orders are
antiferromagnetic (AF) and ferromagnetic (f), respectively
[when spin (orbital) order is vanishing, it is the paramagnetic state 
and is referred to
as P (p) state].

After a simple calculation,  $\Phi_2$ in Eq.(25) becomes 
\begin{equation}
\Phi_2 =  \int d\varepsilon \; f(\varepsilon) \; ( 1/\pi) \;
{\rm Im} \:\:
\sum_{k} \sum_{\sigma} \sum_{\tau} \; 
{\rm ln} \: \left( q_{\tau\sigma}^2  
\left[ X_{A\tau\sigma}(\varepsilon) X_{B\tau\sigma}(\varepsilon) 
- \varepsilon_{k}^2 \right] \right), 
\end{equation}
with the band-narrowing factor, $q_{\tau\sigma}$,  given by
\begin{equation}
q_{\tau\sigma} 
= \surd \overline{r_{A\tau\sigma} r_{B\tau\sigma}}.
\end{equation}
where $r_{A\tau\sigma}$ and $r_{B\tau\sigma}$ 
are given by Eqs.(56)-(60) depending 
on a kind of the spin and orbital orderings.
The expressions for $R_{i\tau\sigma}$ and 
$n_{i\tau\sigma}$ given by Eqs.(47) and (48) become
\begin{equation}
R_{i\tau\sigma} 
= \int d\varepsilon \; f(\varepsilon) \;
\left(-1/\pi \right) \; {\rm Im}
\left[ \left( \Omega_{\tau\sigma}/r_{i\tau\sigma} \right) \;
F_0(\Omega_{\tau\sigma}) \right],
\end{equation}

\noindent
\begin{equation}
n_{i\tau\sigma} 
= \int d\varepsilon \; f(\varepsilon) \;
\rho_{i\tau\sigma}(\varepsilon),
\end{equation}
where $\rho_{i\tau\sigma}$ is the local density of states 
of electrons with spin $\sigma$ in the band $\tau$
on the site $i$, given by
\begin{eqnarray}
\rho_{i\tau\sigma}(\varepsilon) 
& = & (-1/\pi) \; {\rm Im}
\; \left[ K_{A\tau\sigma}(\varepsilon)/r_{A\tau\sigma} \right] 
\;\;\;\;\;\;\;\;\mbox{($i \in A$)}, \nonumber  \\
& = & (-1/\pi) \; {\rm Im}
\; \left[ K_{B\tau\sigma}(\varepsilon)/r_{B\tau\sigma} \right] 
\;\;\;\;\;\;\;\;\mbox{($i \in B$)},
\end{eqnarray}
with
\begin{equation}
K_{A\tau\sigma}(\varepsilon) = \left[ X_{B\tau\sigma}(\varepsilon)
/X_{A\tau\sigma}(\varepsilon) \right]^{1/2}
F_0(\Omega_{\tau\sigma}),
\end{equation}
\begin{equation}
K_{B\tau\sigma}(\varepsilon) = \left[ X_{A\tau\sigma}(\varepsilon)
/X_{B\tau\sigma}(\varepsilon) \right]^{1/2}
F_0(\Omega_{\tau\sigma}),
\end{equation}
\begin{equation}
\Omega_{\tau\sigma}(\varepsilon) 
= \left[ X_{A\tau\sigma}(\varepsilon) \:
X_{B\tau\sigma}(\varepsilon) \right]^{1/2},
\end{equation}
\begin{equation}
F_0(\varepsilon) = \int d\varepsilon' \rho_0(\varepsilon')
/(\varepsilon - \varepsilon'),
\end{equation}
$\rho^0(\varepsilon)$ being the unperturbed density of states.
The optimum values of fields of $\nu$, $\xi$,
$\psi$ and $\eta$, the order  parameters 
of $M$, $O$ and $P$ and
the occupancies, the average of the bosons
of $d_{0\tau}$ {\it et al.}, are self-consistently
determined by Eqs.(11)-(14) with Eqs.(38)-(48).

The expressions for the free energy and the 
self-consistent equations given by Eqs.(36) and (38)-(48) 
without the orbital ordering 
($O = P = 0$ and $\phi = \eta = 0$) reduce to those of the GA for 
the P-p state \cite{Lu94},\cite{Okabe96} 
and to that for the AF-p state \cite{Hasegawa97b}. 
Thus the present slave-boson mean-field theory 
becomes the generalization 
to the cases including both the spin- and orbital 
ordered states in the DHM.

\vspace{0.5cm}
\noindent
{\it 2.3 Comparison with the Hartree-Fock Theory}

It is worth to compare the expression for the free energy
in the present theory given by Eq.(36)
with that in the Hartree-Fock 
approximation given by
\begin{eqnarray}
F_{\rm HF} & = &(1/2) \sum_{i} 
\left[ (\varepsilon_{F} - \nu_i)N_i 
+  \xi_i M_i + \phi_i O_i + \eta_i P_i \right]  \\
& & + \int d\varepsilon \; f(\varepsilon)\;  (-1/\pi) \; 
{\rm Im} \:\:
{\rm Tr} \:\: {\rm ln} \:  G_{HF}(\varepsilon).  \nonumber
\end{eqnarray}
with
\begin{equation}
G_{\rm HF}(\varepsilon) = (\varepsilon - H_{\rm HF})^{-1}
\end{equation}
\begin{equation}
H_{\rm HF} = 
\sum_{i j} \sum_{\sigma} \sum_{\tau} 
t_{ij} 
c^\dagger_{i\tau \sigma} c_{j\tau \sigma}
+ \sum_i \sum_{\sigma} \sum_\tau  
\left[ \nu_{i} - \sigma (h_i + \xi_{i} )
- \tau (g_i + \phi_i )
- \sigma\tau\: \eta_i \right] \:
c^\dagger_{i\tau \sigma} c_{i\tau \sigma},
\end{equation}
$\nu_i$, $\xi_i$, $\phi_i$ and $\eta_i$ being given by
\begin{equation}
\nu_i = (1/4)(U_0 + U_1 + U_2) N_i,
\end{equation}
\begin{equation}
\xi_i = (1/4)(U_0 + U_1 - U_2) M_i,
\end{equation}
\begin{equation}
\phi_i = (1/4)(- U_0 + U_1 + U_2) O_i,
\end{equation}
\begin{equation}
\eta_i = (1/4)(U_0 - U_1 + U_2) P_i.
\end{equation}
Here $N_i$, $M_i$, $O_i$, $P_i$ and $\varepsilon_F$ 
are determined self-consistently by Eqs.(11)-(14) with (73) -(76).
It should be noted that $\nu_i$, $\xi_i$, $\phi_i$ and $\eta_i$
in the Hartree-Fock approximation are given by Eqs.(73)-(76)
while those in the present theory are variationally determined
by Eqs.(38)-(41). 
The free energy of our theory is expected to be generally
lower that the Hartree-Fock free energy, as numerically
shown for the P-p, AF-p and P-af cases.

\begin{center}
{\bf III. CALCULATED RESULTS}
\end{center}

  The formalism presented in the preceding section can be
applied to various types of spin- and/or orbital-ordered 
states.
Since calculated results for P-p and AF-p states have been published
in our previous papers
\cite{Hasegawa97a}-\cite{Hasegawa97c},
we here present some calculations only for orbitally ordered
(P-af) state in order to demonstrate the feasibility of our theory.

   We perform numerical calculations 
for the simple-cubic model with nearest-neighbor hoppings.
Input parameters for our calculations are the non-interacting
density of states,  $\rho_0(\varepsilon)$, 
the Coulomb and exchange interactions, $U$, $J$, and the total 
number of electrons, $N$, which is two for the
half-filled case under consideration.
We employ the approximate, analytic expression 
for  $\rho_0(\varepsilon)$ of the simple-cubic density of 
states \cite{Hasegawa97b}.
%
The energy and the interactions are hereafter measured in units
of a half of its total band width.
The calculated ground-state energy without interactions ($U=J=0$)
is $\varepsilon_0 = -0.3349$, 
which is in good agreement with the exact value of $- 0.3341$
\cite{Yokoyama87}.
Details of the calculation method is explained 
in Ref.\cite{Hasegawa97b}.

\vspace{0.5cm}
\begin{center}
\noindent
{\bf  A. $J=0$ Case }
\end{center}

     We firstly discuss the case of
vanishing exchange interaction ($J = 0$),
for which the P-af state is equivalent to the AF-p state
discussed in Ref.\cite{Hasegawa97b}.
The $U$ dependences of the orbital order $O$, band-narrowing
factor $q$, and the energy $E$ of the P-af state
are given by those of the sublattice magnetization $M$, $q$, and $E$
of the AF-p state shown in Figs.1, 3 and 5, respectively, 
of Ref.\cite{Hasegawa97b}.
The U-dependence of the occupancies of the P-af state
is obtainable from that of the AF-p state shown in Fig.4 
of Ref.\cite{Hasegawa97b};
$d_{0a}=d_{0b}\equiv d_0$,
$d_{1a}=d_{1b}\equiv d_1$,
$d_{a\uparrow}=d_{b_\uparrow}\equiv d_\uparrow$,
$d_{a\downarrow}=d_{b \downarrow}\equiv d_\downarrow$,
$t_{a\uparrow}=t_{b \uparrow}\equiv t_\uparrow$, and
$t_{a\downarrow}=t_{b \downarrow}\equiv t_\downarrow$
in Fig.4 of Ref.\cite{Hasegawa97b}
should be read with the following changes:
$d_{oa} \leftrightarrow d_{2 \uparrow}$, 
$d_{ob} \leftrightarrow d_{2 \downarrow}$, 
$t_{a\downarrow} \rightarrow t_{b \downarrow}$,
$t_{b\downarrow} \rightarrow t_{b \uparrow}$, and
$t_{b\uparrow} \rightarrow t_{a \downarrow}$,
the others being the same for the AF-p and P-af states
({\it a} and {\it b} denote the two orbital bands).

\vspace{0.5cm}
\begin{center}
\noindent
{\bf B. Finite $J$ Case }
\end{center}

     Next we introduce  the exchange interaction $J$ into
our calculation.
In the case of finite $J$, the equivalence between the 
AF-p and P-af states mentioned above for $J=0$ is no more preserved.
Solid curves in Fig. 1 show the $J$ dependence of the orbital
order $O$ for $U$=0.5, 1.0 and 1.5.
For a comparison, results of the HFA are also shown by dashed curves.
With increasing the value of $J$, the orbital order $O$ in the P-af state
decreases.  Because of the large orbital fluctuations,
the orbital order in the GA is much smaller than that in the HFA, and it
almost vanishes for $J > 0.05$ where the P-p state is realized.
This is in contrast with the sublattice magnetization $M$
in the AF-p state, which increase as increasing $J$ \cite{Hasegawa97b}.
This is easily understood because the effective fields of $O$
and $M$ are given by  $(U-5J)/4$ and $(U+J)/4$, respectively, in the HFA 
(Eqs.(2)-(4),(73),(74)).

The densities of states of the P-af state for $J$ = 0.02 and 0.0 with
$U=1.0$ are shown in Figs. 2(a) and 2(b), respectively,
where $\rho_{a\uparrow}=\rho_{a\downarrow}\equiv\rho_a$ and
$\rho_{b\uparrow}=\rho_{b\downarrow}\equiv\rho_b$.
It is noted that Fig. 2(b) for $J=0$ expresses 
also the density of states of the
AF-p state if we read 
$\rho_{a} \rightarrow \rho_{\uparrow} 
(= \rho_{a \uparrow} = \rho_{b \uparrow} )$ and
$\rho_{b} \rightarrow \rho_{\downarrow} 
(= \rho_{a \downarrow} = \rho_{b \downarrow} )$ 
\cite{Hasegawa97b}.
When the $J$ value is increased from 0.0 to 0.02, the orbital
order decreased and the polarization in the densities of states
decreases as shown by Figs. 2(a) and 2(b).

Figure 3 shows the $J$ dependence of the band-narrowing factor $q$
for $U$=0.5, 1.0 and 1.5. With increasing $J$, $q$ increases
slightly for $U=0.5$ while it decreases 
for $U$= 1.0 and 1.5. 

The $J$ dependence of the occupancies
is shown in Fig.4, where from the symmetry, we get
$d_{2\uparrow}=d_{2\downarrow}\equiv d_2$,
$t_{a\uparrow}=t_{a\downarrow}\equiv t_a$ and
$t_{b\uparrow}=t_{b\downarrow}\equiv t_b$.
When $J$ is increased, $d_{ob}$, $d_{1a}$, $d_{1b}$, $t_a$
and $f$ increase whereas $d_{0a}$ and $d_{0b}$ decrease.
At $J>0.05$, the system becomes the P-p state where 
$d_{0a}=d_{0b}$ and $t_a=t_b$ (see Fig. 1).

In order to investigate the relative stability 
between the AF-p and P-af states, we calculate
their total energies $E$, which are shown in Fig. 5.
The bold (thin) solid curve expresses $E$ of the AF-p state 
in the GA (HFA), and the bold (thin) dashed curve $E$ of 
the P-af state in the GA (HFA).
For $J=0$, the AF-p and P-af states have the same total
energy. We note that for all the $J$ values investigated,
$E$ in the GA is always lower than that in the HFA
for the Af-P and p-AF states, as well as for the P-p state 
\cite{Hasegawa97a},\cite{Hasegawa97b}.
For finite (positive) $J$, the energy of the AF-p state
is lower than that of the P-af state both in the
GA and HFA.

\vspace{1.0cm}
\noindent
\begin{center}
{\large\bf IV. CONCLUSION AND DISCUSSION}
\end{center}

     To summarize, we have developed the theory of the spin- and
orbital-ordered states in the DHM,
generalizing our slave-boson mean-field theory \cite{Hasegawa97a}.
In order to demonstrate the feasibility of our theory,
we have presented numerical calculations of the antiferromagnetic
orbital (P-af) state,
showing the $J$ dependences of the orbital order $O$, the band narrowing
factor $q$ and the occupancies.  From the calculation of 
the total energy $E$, the antiferromagnetic orbital (P-af)
state is shown to be unstable against the 
antiferromagnetic spin (AF-p) state except for $J=0$ 
for which the both states are energetically degenerate.

   In order to more investigate the role of the exchange 
interaction on orbital order in the DHM, 
we have made numerical calculations also for the {\it negative} $J$, 
although $J$ is conventionally taken to be positive.
We notice in Fig.1 that for the negative exchange interaction, 
the orbital order $O$ is
considerably increased: we cannot obtain the solution
at $J < 0.048$ ($J < 0.014$) for $U=1.0$ ($U=1.5$).
This increase in $O$ is induced by a significant increase
in $d_{0a}$, as shown in Fig. 4.
As was discussed in Ref.\cite{Hasegawa97b}, 
the negative exchange interaction
is not favorable for the sublattice magnetization $M$
in the AF-p state.  Then the increase in $O$ and the 
decrease in $M$ for the negative $J$ lead to the P-af state
to be more stable than the AF-p state,
as the energy calculation shows in Fig. 5.

We previously showed from a comparison between the 
calculations using the GA \cite{Hasegawa90} and 
the variational Monte-Calro method 
\cite{Yokoyama87}
that the GA is a fairly good approximation for the 
one-, two- and three-dimensional simple SHM 
(the GA is exact for infinite dimensional SHM). 
We expect that this would hold also for the DHM although
no variational calculations for the DHM have been
reported yet. 
We can apply our theory to states in which both spin and orbital
orderings coexit such as the AF-f and F-f states.
A study of the N-U phase diagram 
(N: electron number, U: interaction) of the simple-cubic DHM,
as was made within the HFA \cite{Inagaki}, will be interesting. 
It would be also promising to investigate the temperature-interaction
phase diagram of the DHM, by generalizing our approach 
\cite{Hasegawa89}\cite{Hasegawa90b}
in which the effects of electron correlation and 
thermal spin fluctuations are properly taken 
into account.

\section*{Acknowledgements}
It is my great pleasure to dedicate the present paper to 
Dr. Martin Gutzwiller on the occasion of his 75th birthday.
This work has been supported partially by
a Grant-in-Aid for Scientific Researach (B) 
from the Ministry of Education, Science, Sports and Culture of Japan.

\begin{figure}
\caption{
The $J$ dependence of the orbital order $O$ in the P-af state
calculated by the GA
(solid curves) and HFA (dashed curves) 
}
\label{fig1}
\end{figure}

\begin{figure}
\caption{
The local densities of state in the P-af state
for (a) $J=0.02$ and (b) $J=0.0$ with $U=1.0$:
$\rho_{a\uparrow}=\rho_{a\downarrow}\equiv \rho_a$ (solid curves) and
$\rho_{b\uparrow}=\rho_{b\downarrow}\equiv \rho_a$ (dashed curves).
}
\label{fig2}
\end{figure}

\begin{figure}
\caption{
The $J$ dependence of the band-narrowing factor $q$
in the P-af state.
}
\label{fig3}
\end{figure}

\begin{figure}
\caption{
The $J$ dependence of the occupancies in the P-af state:
$d_{2\uparrow}=d_{2\downarrow}\equiv d_2$,
$t_{a\uparrow}=t_{a\downarrow}\equiv t_a$ and
$t_{b\uparrow}=t_{b\downarrow}\equiv t_b$.
}
\label{fig4}
\end{figure}

\begin{figure}
\caption{
The $J$ dependence of the total energies $E$ 
of the AF-p state in the GA (bold, solid curve)
and HFA (thin, solid curve), and $E$ of the
P-af state in the GA (bold, dashed curve)
and HFA (thin, dashed curve).
}
\label{fig5}
\end{figure}


\begin{references}
 

\bibitem{Gutzwiller63}M. C. Gutzwiller, Phys. Rev. Lett. {\bf 10}, 159 (1963);
Phys. Rev. {\bf 137}, A1726 (1965).
\bibitem{Kaplan82}T. A. Kaplan, P. Horsch and P. Fulde,
Phys. Rev. Lett. {\bf 49}, 889  (1982).

\bibitem{Yokoyama87}H. Yokoyama and H. Shiba, 
J. Phys. Soc. Jpn.  {\bf 56}, 1490 (1987);
{\it ibd} {\bf 56}, 3582 (1987).


\bibitem{Brinkman70}W. F. Brinkman and T. M. Rice, 
Phys. Rev. B {\bf 2}, 4302 (1970).

\bibitem{Gebhard97}F. Gebhard, 'The Mott Metal-Insulator Transition-
Models and Mehtods', Springer Tracts in Modern Physcs, Vol. 137
(Springer, Berlin 1997).



\bibitem{Chao71}K. A. Chao and M. C. Gutzwiller,
Phys. Rev. B {\bf 4}, 4034 (1971).
\bibitem{Lu94}J. P. Lu,  Phys. Rev. B {\bf 49}, 5687 (1994).
\bibitem{Okabe96} T. Okabe, J. Phys. Soc. Jpn. {\bf 65}, 1056  (1996).
\bibitem{Bunemann97}J. B\"{u}nemann and W. Weber, 
Phys. Rev. B {\bf 55}, R4011 (1997).
\bibitem{Bunemann98}J. B\"{u}nemann, W. Weber, and F. Gebhard, 
Phys. Rev. B {\bf 57}, 6896 (1998).

\bibitem{Hasegawa97a}H. Hasegawa, J. Phys. Soc. Jpn. {\bf 66}, 1391 (1997).
\bibitem{Hasegawa97b}H. Hasegawa, 
Phys. Rev. B {\bf 56}, 1196 (1997).
\bibitem{Hasegawa97c}H. Hasegawa, J. Phys. Soc. Jpn. {\bf 66}, 3522 (1997). 
\bibitem{Fresard97}R. Fr\'{e}sard and G. Kotliar, 
Phys. Rev. B {\bf 56}, 12909 (1997).

\bibitem{Kotliar96}G. Kotliar and H. Kajueter, Phys. Rev. B {\bf 54}, 
R14221 (1996).
\bibitem{Rozenberg96}M. J. Rozenberg, Phys. Rev. B {\bf 55} R4855 (1997).





\bibitem{Gill87}W. Gill and D. J. Scalapino, 
Phys. Rev. B {\bf 35}, 215 (1987).

\bibitem{Bunemann90}J. E. Han, M. Jarrel, and D. L. Cox, 
Phys. Rev. B {\bf 58}, 14a99 (1998).

\bibitem{Motome98}Y. Motome and M. Imada,
J. Phys. Jpn. {\bf 67}, 3199 (1998).



\bibitem{Dorin93}V. Dorin and P. Schlottmann,
Phys. Rev. {\bf 47}, 5095 (1993).
\bibitem{Kotliar86}G. Kotliar and A. E. Ruckenstein,
Phys. Rev. Lett. {\bf 57}, 1362 (1986).

\bibitem{Hasegawa89}H. Hasegawa, J. Phys. Cond. Matter.  
{\bf 1}, 9325 (1989);

\bibitem{Hasegawa90}H. Hasegawa, 
Phys. Rev. B {\bf 41}, 9168 (1990).

\bibitem{Inagaki}S. Inagaki and R. Kubo,
Int. J. Mag. {\bf 4}, 139 (1973).

\bibitem{Hasegawa90b}H. Hasegawa, 
Prog. Theor. Phys. Suppl.{\bf 101}, 463 (1990).










\end{references}
\end{document}